\documentstyle[12pt] {article}

\newcommand{\sect}[1]{ \section{#1} \setcounter{equation}{0} }

\newcommand{\req}[1]{(\ref{#1})}

\newcommand{\f}{\hat{f}}

\newcommand{\nwc}{\newcommand}
\nwc{\hyp} {\hyphenation}
\nwc{\be}  {\begin{equation}}
\nwc{\ee}  {\end{equation}}
\nwc{\ba}  {\begin{array}}
\nwc{\ea}  {\end{array}}
\nwc{\bdm} {\begin{displaymath}}
\nwc{\edm} {\end{displaymath}}

\nwc{\bea} {\be\ba{lcl}}
\nwc{\eea} {\ea\ee}

\nwc{\bda} {\bdm\ba{lcl}}
\nwc{\eda} {\ea\edm}

\nwc{\bc}  {\begin{center}}
\nwc{\ec}  {\end{center}}

\nwc{\ds}  {\displaystyle}
\nwc{\bmat}{\left(\ba}
\nwc{\emat}{\ea\right)}
\nwc{\nn} {\nonumber}
\nwc{\nnn} {\nonumber \vspace{.2cm} \\ }
\nwc{\ra}{\rightarrow}
\nwc{\lra}{\longrightarrow}

\nwc{\p} {\partial}
\nwc{\SS} {S}

\nwc{\sieb}{\bf \overline{27}}

\nwc{\scr}  {\scriptstyle}
\nwc{\tx}  {\textstyle}
\nwc{\scs} {\scriptscriptstyle}

\nwc{\ov}  {\overline}

\nwc{\hb}  {\bar h}
\nwc{\xb}  {\bar x}
\nwc{\yb}  {\bar y}
\nwc{\zb}  {\bar z}
\nwc{\wb}  {\bar w}
\nwc{\Ob}  {\bar O}
\nwc{\Yb}  {\bar Y}

\nwc{\e} {\epsilon}
\nwc{\de} {\delta}
\nwc{\Th} {\Theta}
\nwc{\th} {\theta}
\nwc{\al} {\alpha}
\nwc{\si} {\sigma}
\nwc{\om} {\omega}
\nwc{\Om} {\Omega}
\nwc{\Ga} {\Gamma}

\nwc{\sipk} {\si^{+k}}
\nwc{\simk} {\si^{-k}}
\nwc{\sipl} {\si^{+l}}
\nwc{\siml} {\si^{-l}}

\nwc{\Sc}  {{\cal S}}
\nwc{\Rc}  {{\cal R}}
\nwc{\Dc}  {{\cal D}}
\nwc{\Oc}  {{\cal O}}
\nwc{\Cc}  {{\cal C}}
\nwc{\gc}  {{\cal g}}

\nwc{\Of}  {{\cal O}_f}
\nwc{\Oft} {{\cal O}_{f_2}}
\nwc{\Ofo} {{\cal O}_{f_1}}

\nwc{\Pc}  {{\cal P}}
\nwc{\Mc}  {{\cal M}}
\nwc{\Ec}  {{\cal E}}
\nwc{\Fc}  {{\cal F}}
\nwc{\Hc}  {{\cal H}}
\nwc{\Kc}  {{\cal K}}
\nwc{\Wc}  {{\cal W}}

\nwc{\Fcp} {{\cal F}^\pr}
\nwc{\Hcp} {{\cal H}^\pr}

\nwc{\Xc}  {{\cal X}}
\nwc{\Gc}  {{\cal G}}
\nwc{\Zc}  {{\cal Z}}
\nwc{\Nc}  {{\cal N}}

\nwc{\xc}  {{\cal x}}
\nwc{\Ac}  {{\cal A}}
\nwc{\Bc}  {{\cal B}}
\nwc{\Uc} {{\cal U}}
\nwc{\Vc} {{\cal V}}
\nwc{\Lc} {{\cal L}}
\nwc{\Qc} {{\cal Q}}

\nwc{\lng} {\langle}
\nwc{\rng} {\rangle}

\nwc{\lf} {\left}
\nwc{\ri} {\right}

\nwc{\pr} {\prime}
\nwc{\diag} {{\rm diag}}
\nwc{\inv}  {{\rm inv}}
\nwc{\mod}  {{\rm mod}}
\nwc{\tr}  {{\rm tr}}
\nwc{\im}  {{\rm Im}}
\nwc{\re}  {{\rm Re}}

\nwc{\h} {\frac{1}{2}}
\nwc{\fc} {\frac}

\def\KK{\relax{\rm I\kern-.18em K}}
\def\RR{\relax{\rm I\kern-.18em R}}
\def\NN{\relax{\rm I\kern-.18em N}}
\def\PP{\relax{\rm I\kern-.18em P}}

\def\ZZ{\relax{\sf Z\kern-.4em Z}}
\def\ZZZ{Z\kern -0.31em Z}
\def\QQ{{\rm \kern .25em
             \vrule height1.4ex depth-.12ex width.06em\kern-.31em Q}}
\def\CC{{\rm \kern .25em
             \vrule height1.4ex depth-.12ex width.06em\kern-.31em C}}
\newcommand{\Z}{\ZZ}
\nwc{\z} {$Z_N \times Z_M\ $}
\newcommand{\Rr}{{\rm R}}
\newcommand{\Lr}{{\rm L}}

\if@twoside\oddsidemargin25mm
\else\oddsidemargin25mm\evensidemargin25mm\marginparwidth25mm\fi%
\begin{document}

\begin{titlepage}
\begin{flushright}TUM--TH--151/92 \\ hep-th/9211027 \\ \ \\
October 1992
\end{flushright}
\vfill

\begin{center}
{\large\bf Moduli and Twisted Sector Dependence \\
of \z Orbifold Couplings $^\ast$}   \\
\vskip 1.2cm
{\bf S. Stieberger} \\
\vskip .5cm

{\em Physik Department} \\
{\em Technische Universit\"at M\"unchen} \\
{\em D--8046 Garching, FRG}
\end{center}
\vfill

\vspace{4cm}

\begin{center}
{\bf ABSTRACT}
\end{center}

\begin{quote}
We derive the four point correlation function involving four twist fields for
arbitrary even dimensional  \z orbifold compactifications. Using techniques
from the
conformal field theory the three point correlation functions with
twist fields are determined. Both the choice of the modular background
(compatible with the twists) and of the (higher) twisted sectors involved are
fully general. Our results turn out to be target space duality invariant.
\vskip 5mm \vskip0.5cm
\hrule width 5.cm \vskip 1.mm
{\small\small
\noindent $^\ast$ Supported by the Deutsche Forschungsgemeinschaft}
\normalsize
\end{quote}

%\begin{flushleft}
%October 1992
%\end{flushleft}

\end{titlepage}

\sect{Introduction}

Toroidal Orbifolds \cite{DHVW,IMNQ,FIQ} constitute a wide class of soluble
string
compactifications, which have good phenomenological perspectives.
Furthermore, in this class of models all quantities of interest may be
computed as functions of moduli deformations \cite{DVV,DS}. This makes it
possible to
relate some of these compactifications to $N=2$ {\em Landau--Ginzburg
models} (LGMs) \cite{LVW}: Their massless spectrum and Yukawa couplings can be
compared. These connections are constructed explicitly for the $Z_3 \times Z_3$
orbifold in \cite{CLMN} whereas for other models similar relations are done
only at their critical
points. At the critical points LGMs become the {\em Gepner
models} \cite{G}.
Lots of them are known to agree with their massless spectra, couplings and
discrete
symmetries with \z orbifold constructions at their critical points
\cite{FIQ2,CLN,CK}. The equivalence for arbitrary values of the moduli follows
then by
applying marginal deformations.
To investigate these relations in more detail it is very important to know the
Yukawa couplings to be able to make such a comparison.

Until now $Z_N$ orbifolds received most attention, but in fact the \z
orbifolds are also very interesting,
because they are the most general Abelian orbifolds leaving four
uncompactified dimensions and
one unbroken supersymmetry \cite{FIQ}. \z orbifolds also give rise to
phenomenologically attractive string vacua \cite{IL}. Their couplings can be
used to
make detailed phenomenological studies like calculations of quark and lepton
masses and
mixing angles.
In string theory  these couplings receive moduli dependent instanton
corrections \cite{DFMS}. These non--perturbative effects lead to exponentially
suppressed factors that depend on the instanton action. This last property
naturally yields hierarchies of Yukawa couplings \cite{DFMS}.

Recently the Yukawa couplings involving three twist fields were calculated for
the $Z_3 \times Z_3$\ orbifold in a particular region of moduli space. In
the language of twisted Narain compactifications this region is determined
by setting the antisymmetric tensor background
field \cite{BLS} to zero. A calculation of all bosonic \z orbifold Yukawa
couplings with complete moduli dependence had not been done until now.

Whereas the $Z_3 \times Z_3$ couplings can be calculated in the spirit of
\cite{DFMS,LMN2,EJLM},
for the \z orbifold couplings in general higher order twist fields appear.
The calculation of couplings involving higher order twist fields was initiated
in \cite{BKM} and completed in \cite{sjls}.
Apart from this, there are several new aspects to take into account for
orbifolds with two independent
generators of the point group.
It turns out to be useful to split the whole target space into two
dimensional complex subspaces and to calculate the couplings for all these
subspaces independently. This way the calculations can be reduced to $Z_M$
orbifold
calculations. For this splitting it is necessary to have a block--diagonal
antisymmetric
background field. This not a restriction, at least for six dimensions,
because none of the  \z
orbifolds in $d=6$ allow any other structure \cite{s}. It is also important
to realize that fixed planes appear
already in the generating twists. In this short note we show how the
subtleties involved can be overcome and we give results for all the
couplings in these models.

The paper is organized as follows:
In the second section we present the basic properties of \z
orbifolds and derive the general four twist correlation function.
We allow a
general target space metric and antisymmetric tensor field compatible with the
twisting. In section three we use operator product expansions for the twist
fields  to factorize the four point correlation function and we calculate
all three
point couplings.
In the last section we define a large class of modular duality
transformations and
show that all the calculated couplings are invariant under this modular
group, provided that the twisted sector ground states are transformed
appropriately.

\sect{The four point function}

The class of \z orbifold models considered here may be described by the action
\cite{NSW}:

\be
S(G,B) = \frac{1}{2 \pi} \int  dz d \bar{z} \ \bar{\p} X^j (G+B)_{jk}\p X^k\ .
\label{action}
\ee
The coordinate fields $X^j(z,\bar z)$
describe the embedding of the string in
the $d$--di\-men\-sio\-nal compact target space, where $d$ is assumed to be
even. The constant
target space metric $G$ and the antisymmetric background tensor $B$
refer to the coordinate basis. As discussed in
\cite{GRV,EJLM} $G$ can be
set to ${1\over2} {\bf 1}_d$ via a coordinate
transformation.

The point group of a \z orbifold has two independent generators denoted by
$\Th$ and $\Omega$ (these will sometimes be referred to as the twists). The
point group then is

$$\{\Th^{k^\Th} \Om^{k^\Om}\ |\  0 \leq k^\Th < N\ ,\ 0 \leq k^\Om < M\} \ .$$

The eigenvalues of the generators are the phases $\th_j=e^{2 \pi i
\al_j^\Th}\ ,\ \om_j=e^{2 \pi i
\al_j^\Om}\ ,\ j = 1, \ldots , d/2$ respectively (and their complex
conjugates) such
that $N \al_j^\Th, M \al_j^\Om \in \Z$. The twists are orthogonal with respect
to the metric $G$.
The fields $X$ are subject to the twisted boundary conditions:
\bea
 X(e^{2\pi i}z,e^{-2\pi i}\bar z)=(\Th X)(z,\bar z) + 2\pi w\nnn
X(e^{2\pi i}z,e^{-2\pi i}\bar z)=(\Om X)(z,\bar z) + 2\pi w'\ .
\label{bocon}
\eea
The translation vectors $w,w'$ belong to the lattice which describes the
compact
$d$ dimensional toroidal target space. From the theory of Abelian finite groups
we know that $M$
must be
divisible by $N$ \cite{FIQ}. Additional restrictions arise because both twists
must be simultaneously automorphisms of the
lattice $\Lambda$.

The embedding $X$ is split into a classical {\em instanton} solution
and a quantum fluctuation part \cite{DFMS}:

\be
X^j(z,\bar z)=X_{\rm cl}^j(z,\bar z)+X_{\rm qu}^j(z,\bar z)\ ,
\ \ \ j=1, \ldots,d\ ,
\ee
where $X_{\rm cl}^j$ denotes a solution to the classical equation of motion
$\p \bar \p X^j(z,\bar z)=0$ following from \req{action}. $X_{\rm qu}^j$
describes the quantum fluctuations around these instantons.

Before we define the twist fields we have to realize how the group elements act
on the $d$ dimensional target space.
We chose the twists in their simplest $2 \times 2$ block
diagonal form. If we then pass to new complex coordinate fields

\be
Y^j(z,\bar z) := X^{2j-1}(z,\bar z) + i X^{2j}(z,\bar z)\ ,
\qquad j = 1,\ldots,d/2\ ,
\ee
the twist operations become $Y^j \mapsto \th_j Y^j$ and $Y^j \mapsto \om_j
Y^j$ respectively.
The action of the twist $\Th^{k^\Th}\Om^{k^\Om}$ on the $j$th complex subspace
can be represented in the form
\be
\th_j^{k^\Th} \om_j^{k^\Om} \equiv e^{2 \pi i (k_j^{\Th}+k_j^{\Om})}=:e^{2
\pi i k_j}\ ,\ k_j:=[k_j^\Th+k_j^\Om]=[k^\Th\al_j^\Th+k^\Om\al_j^\Om] \in
[0,1)\ .
\label{Mor}
\ee
Since $M$ is divisible by $N$ this identity enables us to interpret the action
of the twist as a
twist of a proper $Z_M$ orbifold.

A twist field is characterized by the sector $(k^\Th,k^\Om)$ whose vacuum
it creates and a fixed
point $f$ satisfying $\Th^{k^\Th}
\Om^{k^\Om} f=f +w\ ,\ w \in \Lambda$ .
The twist field is defined by its operator product expansion (OPE) with the
coordinate
differentials e. g.:

%\be \begin{array}{clccr}
\be
 \ds{\p Y^j(z,\bar z)\  \sigma^{k^\Th
\hspace{-.22cm},\hspace{.09cm}k^\Om}_{f}(w,\bar w)}
% & = &
=
\ds{(z-w)^{-(1-k^\Th_j-k^\Om_j)}
 }
%&
\ds{ (\tau_j)^{k^\Th \hspace{-.22cm},\hspace{.09cm}k^\Om}_f(w,\bar w)}+\ldots
% & \nnn
% \ds{\p \bar Y^j(z,\bar z)\
% \sigma_{f}^{k^\Th \hspace{-.22cm},\hspace{.09cm}k^\Om}(w,\bar w)} & = & \ds{(
% z-w)^{-k^\Th_j-k^\Om_j}
% }&\ds{ (\tau'_j)^{k^\Th \hspace{-.22cm},\hspace{.09cm}k^\Om}_f(w,\bar
%%w)}+\ld% ots & \nnn
% \ds{\bar \p
% Y^j(z,\bar z)\  \sigma_{f}^{k^\Th \hspace{-.22cm},\hspace{.09cm}k^\Om}(w,\bar
% w)} & = & \ds{(\bar z-\bar
%  w)^{-k^\Th_j-k^\Om_j}}&\ds{ (\tilde{\tau}_j^\prime )^{k^\Th
%  \hspace{-.22cm},\ hspace{.09cm}k^\Om}_f(w,\bar w)}+\ldots & \nnn
% \ds{\bar \p \bar Y^j(z,\bar z)\  \sigma_{f}^{k^\Th \hspace{-.22cm},\hspace{.0
%  9 cm}k^\Om}(w,\bar w)}
% & = & \ds{(\bar z-\bar w)^{-(1-k^\Th_j-k^\Om_j)}}
% &\ds{(\tilde{\tau}_j)^{k^\Th \hspace{-.22cm},\hspace{.09cm}k^\Om}_f(w,\bar
% w) }+\ldots \; , &
% \end{array}
\label{twistOPE}
\ee
Similar expressions may be found for $\p \bar Y,\ \bar \p Y,\ \bar \p \bar Y$.
On the right hand side of (\ref{twistOPE}) a so called {\em excited twist
field} is introduced \cite{ejss}.

The adjoint of
$\sigma^{k^\Th \hspace{-.22cm},\hspace{.09cm}k^\Om}_{f}$ is
$\sigma^{-k^\Th \hspace{-.22cm},\hspace{.11cm}-k^\Om}_{f}$
and their conformal weights are given by

\be
 h_{(k^\Th \hspace{-.22cm},\hspace{.09cm}k^\Om)}^\si=\bar h_{(k^\Th
\hspace{-.22cm},\hspace{.09cm}k^\Om)}^\si =\sum_{j=1}^{d/2}h_{k_j}^\si  =
 \h \sum_{j=1}^{d/2} k_j(1-k_j)\ .
 \label{hkj}
\ee
This immediately follows due to the $Z_M$ representation \req{Mor} of $
\Th^{k^\Th}\Om^{k^\Om}$ and \cite{ejss}.
These twist fields are in general not twist invariant and are therefore not
physical fields. One
has to construct twist invariant linear combinations. Since the physical twist
fields are linear combinations of the usual, we may for
simplicity only consider the latter in the following. All the results are
easily extended to couplings involving the twist invariant operators.

The twist field $\si^{k^\Th \hspace{-.22cm},\hspace{.09cm}k^\Om}_{f}(0,0)$
imposes the
boundary condition
\be
 X_{\rm cl}(e^{2\pi i}z,e^{-2\pi i}\bar z)=(\Th^{k^\Th} \Om^{k^\Om} X_{\rm
cl})(z,\bar z) + 2\pi w
\label{bocon}
\ee
on the classical field and also on the full field. Therefore the quantum part
only feels the {\em local} monodromy:

\be
X_{\rm qu}(e^{2\pi i}z,e^{-2\pi i}\bar z)=(\Th^{k^\Th} \Om^{k^\Om}
X_{\rm qu})(z,\bar z)\ .
\ee
In general the instanton solutions of the equations of motion $Y_{\rm cl}$
receive branch point singularities
as a consequence of the twisted boundary conditions \req{bocon}. E. g. $\p
Y^j(e^{2\pi i} z) = e^{2 \pi ik_j} \p Y^j(z)$.

In \cite{EJLM} it is shown that \req{action} is only consistent with
\req{bocon} if one requires for the antisymmetric tensor field $B$\ that
$[B, \Th]
=0$ {\em and\ } $[B,\Om]=0.$
These equations put strong restrictions on the number of possible independent
parameters of the $B$
field \cite{MS1}. For the six dimensional \z orbifold compactifications
\cite{FIQ} the
twists $\Th, \Om$ leave fixed always at least one plane. By this fact one gets
the block diagonal structure of $B$
which is necessary for the factorization of the theory into three {\em
independent} complex subspaces \cite{s}. Therefore one has only
three untwisted $(1,1)$--moduli in six dimensions. For that reason we assume a
block diagonal structure of $B$ in $d$ dimensions, too. One could also consider
{\em discrete
torsion} \cite{FIQ} but for these cases the action \req{action} is no longer
appropriate.
\\

We begin by calculating the four twist correlation function

\be
Z_{\{f_{i}\}}(x,\bar x)=\lim_{|z_{\infty}| \ra
\infty}|z_{\infty}|^{4h^\si_{(l^\Th \hspace{-.22cm},\hspace{.09cm}l^\Om)}}\
\lng
\sigma_{f_1}^{-k^\Th\hspace{-.22cm},\hspace{.11cm}-k^\Om}(0,0)\sigma_{f_2}^{k^\Th \hspace{-.22cm},\hspace{.11cm}k^\Om}(x,\bar
x)\sigma_{f_3}^{-l^\Th\hspace{-.22cm},\hspace{.11cm}-l^\Om}(1,1)\sigma_{f_4}^{l^\Th \hspace{-.22cm},\hspace{.11cm}l^\Om}(z_{\infty},\bar z_{\infty}) \rng
\label{4pf}
\ee
which will be used to derive various three point functions by factorization.
Our calculations are based on the path integral formalism \cite{DFMS}.
Since we are only
interested in their moduli dependence we dropped the fermionic part. Note that
from this part additional selection rules \cite{KO} arise.

The action is
bilinear in the fields $X^j$, therefore the path integral representation of
the twisted vacuum configuration \req{4pf} splits into two factors, one coming
from the {\em classical} instanton solutions $X_{\rm cl}(z,\bar z)$ and the
other
describing the {\em quantum fluctuations} around them \cite{DFMS}.
%\be
%Z_{\{f_{i}\}}(x,\bar x) = Z_{\rm qu}(x,\bar x)\ \sum_{X_{\rm cl}}
%e^{-S [\h{\bf 1}_d,B;X_{\rm cl}]}\ .
%\label{z}
%\ee
The calculation can be performed {\em independently} in every subspace
\cite{s} due
to the block--diagonal $B$ field. To proceed we make use of the $Z_M$
orbifold structure \req{Mor} in every
complex subspace. We choose an auxiliary orbifold to calculate the part of
the
correlation function which corresponds to the $j$th two dimensional
subspace. For
$Mk_j,Ml_j$ we define their greatest common divisor $\phi_j$
and $Mk_j=\phi_j k_{oj},\ Ml_j=\phi_j l_{oj}$. The corresponding twist
$\Xi$ of
the $Z_M$ orbifold is defined to have the complex phases  $\xi_j=e^{2 \pi i
\frac{\phi_j}{M}}\ ,\ j=1,\ldots,d/2$.
Now we calculate, for every subspace, four point correlation
functions with twist orders $k_{oj}, l_{oj}$ associated to the $Z_M$ orbifold.
This is already
done
%\footnote{We get different encircling
%numbers for every subspace in these cases. Note also that the greatest common
%divisor of $k_{0j}$ and $l_{0j}$ is $1$. For details see \cite{ejss}.}
in \cite{sjls}.
For the case of fixed planes e. g. $k_j=0$ ($\varphi_j=1$)
we will not get any contribution from these subspaces\footnote{This projection
we will denote by a $\bot$
and $j
\neq \Im$, respectively.} in the classical action
if we restrict us here to a two twist field coupling instead of a four point
coupling.
The problem with fixed tori is treated in
\cite{dirk,ejss}.

After determining the space group selection \cite{ejss} rule for every
subspace
one may put them together in one equation:

\be
(1-\Th^{k^\Th}\Om^{k^\Om}) f_{21}+(1-\Th^{l^\Th}\Om^{l^\Om})f_{43} =
(1-\Th^{k^\Th}\Om^{k^\Om})T_1+(1-\Th^{l^\Th}\Om^{l^\Om})T_2\ ,
\label{srule}
\ee
with $f_{ab}:=f_a-f_b$ and $T_1, T_2 \in \Lambda$.
For the final result \req{4pf} we get

\be
\hspace{-.25cm} Z_{\{f_{i}\}}(x,\bar x) =(2
\pi)^{\frac{d}{2}}|x|^{-4h^\si_{(k^\Th \hspace{-.22cm},\hspace{.09cm}k^\Om)}}
\left[\prod_{j=1 \atop j \neq \Im}^{d/2}\frac{(1-x)^{-l_j (1-k_j )}
(1-\xb )^{-k_j (1-l_j )}}{I_j (x,\bar x )}\right] \hspace{-1.5cm}
\ds{\sum_{\hspace{1.5cm} v \in (1-\Th^{k^\Th}\Om^{k^\Om}) (f_{21} +
\tilde{\Lambda})_{\bot}
    \atop \hspace{.4cm} u \in
(f_{23} + \Lambda)_{\bot}} \hspace{-1cm}
e^{-S [G,B;v,u]}}\ .
\ee
Here $V_{\Lambda}$ is the volume of a unit cell of $\Lambda$ and
$\tilde{\Lambda}=-T_1+\frac{1-\Th^{l^\Th}\Om^{l^\Om}}{1-\Xi}\Lambda$. The
classical action
evaluated for the instanton solutions is  denoted by
$S[G,B; v,u]$,
since it depends on the vectors $v,u$ and on $G\ ,B$. For details see
\cite{sjls,ejss}.
Note that the summation range does not depend on the particular choice of $T_1$
and $T_2$ as long \req{srule} is fulfilled \cite{sjls}.

The function $I_j (x,\bar x )$ is given by
\be
I_j(x,\bar x ):=  % \frac{\sin (pk_j\pi )}{\pi}
     J_{2j}\ov{G}_{1j}(\bar x )H_{2j}(1-x)+
     J_{1j}G_{2j}(x)\ov{H}_{1j}(1-\bar x ) \ ,
\ee
with (the hypergeometric function is abbreviated by $F(a,b;c;x)$)

\be
\ba{rcl}
 \ds{G_{1j}(x):= F(k_j,1- l_j;1;x)}&,&
 \ds{G_{2j}(x):= F(1-k_j,l_j;1;x)} \nnn
 \ds{H_{1j}(x):= F(k_j,1-l_j;1+k_j-l_j;x) }&,&
 \ds{H_{2j}(x):= F(1-k_j,l_j;1+l_j-k_j;x) } \nnn
 \ds{J_{1j}:=\frac{\Ga(k_j) \Ga(1-l_j)}{\Ga(1+k_j-l_j)}}&,&
 \ds{ J_{2j}:=\frac{\Ga(1-k_j)\Ga(l_j)}{\Ga(1+l_j-k_j)}}\ .
\ea
\ee

\sect{The three point functions}

There are two kinds of three point functions involving twist fields: the
coupling of two twisting states to an untwisted sector state, and the coupling
of three twisted states.

The coupling of two twisted sector states to an untwisted one is defined as

\be
C^{(k^\Th
\hspace{-.22cm},\hspace{.09cm}k^\Om)}_{f_2,f_1;p,w}:=\lim_{|z_{\infty}|\ra
\infty} |z_{\infty}|^{4h^\si_{(k^\Th \hspace{-.22cm},\hspace{.09cm}k^\Om)}}  \,
    \lng
V_{-p,-w}(0,0)\si_{f_1}^{-k^\Th\hspace{-.22cm},\hspace{.11cm}-k^\Om}(1,1)
    \si_{f_2}^{k^\Th
\hspace{-.22cm},\hspace{.09cm}k^\Om}(z_{\infty},\zb_{\infty}) \rng \ .
\label{ssv}
\ee
Here $V_{p, w}$ denotes the vertex operator \cite{LMN2,EJLM}

\be
   V_{p,w}=e^{i P_L^t X_L(z)
   +i P_R^t X_R(\bar{z})}
   \label{vertexoperator}
\ee
and
$(h,\bar h)$ its conformal dimension
\bdm
\ba{lclccrcclccr}
    & \ds{h} &=& \ds{\frac{1}{2}P_\Rr^t P_\Rr} &,&
    \ds{\bar h} &=& \ds{\frac{1}{2} P_\Lr^t P_\Lr\ } \\[5mm]
    {\rm with} & \ds{P_\Rr} &=& \ds{p-Bw +{\ds w\over 2}}
    &,& \ds{P_\Lr} &=& \ds{p-Bw-{\ds w\over 2}} \ .
\ea
\edm
The coupling can be obtained by  the four point function using the
general form of the OPE \cite{sjls}. Again we split the calculation into $d/2$
similar problems where we determine
the annihilation couplings for the $Z_M$ orbifold defined in the last chapter
\cite{ejss}. After straightforward
calculations we get

\be
   C_{f_2,f_1;p,w}^{(k^\Th \hspace{-.22cm},\hspace{.09cm}k^\Om)} =
   \lf[\prod_{j=1 \atop j \neq \Im}^{d/2} \de_{k_j}^{-\h(h^j+\bar h^j)} \ri]
   e^{-2 \pi i f_2^t p^t_\bot}
   e^{i \pi p^t_\bot \frac{1}{1-\Th^{k^\Th}\Om^{k^\Om}}w}\
   \delta_{w\in (1-\Th^{k^\Th}\Om^{k^\Om})(f_{21}+\Lambda)_\bot}\ ,
   \label{svs}
\ee
where $h^j, \bar h^j$ are the parts of the conformal dimension of the vertex
operator arising from the $j$th subspace and $$\ln \de_{k_j} \equiv2
\psi(1)-\psi(k_j)-\psi(1-k_j)\ \ ,\ \
\psi(x) \equiv \frac{d \ln\Ga(x)}{dx}\ .$$

%\bdm
%   g_{(k^\Th \hspace{-.22cm},\hspace{.09cm}k^\Om)}(P_\Lr,P_\Rr) = \lf[
%%\frac{1}{M}
%   \ri]^{\frac{1}{2}(P_\Rr^2+P_\Lr^2)}
%   \; \prod_{n=1}^{M-1}|1-e^{2 \pi i n/M}|^{\h(P_\Rr^t
%   \Th^{nk^\Th}\Om^{nk^\Om} P_\Rr+P_\Lr^t \Th^{nk^\Th} \Om^{nk^\Om}P_\Lr)}\ .
%\edm

The Yukawa couplings are given by

\be
Y_{f_a,f_b,f_c}^{(k^\Th
\hspace{-.22cm},\hspace{.09cm}k^\Om\hspace{-.22cm},\hspace{.13cm}l^\Th
\hspace{-.22cm},\hspace{.09cm}l^\Om)}\equiv \lim_{|x|\ra \infty}
     |x|^{4h^\si_{(k^\Th \hspace{-.22cm},\hspace{.09cm}k^\Om)}}
     \lng \si_{f_a}^{k^\Th \hspace{-.22cm},\hspace{.09cm}k^\Om}(x,\bar
x)\si_{f_b}^{l^\Th \hspace{-.22cm},\hspace{.11cm}l^\Om}(1,1)
\si_{f_c}^{-(k^\Th\hspace{-.14cm}+l^\Th\hspace{-.05cm}),-(k^\Om\hspace{-.14cm}+l^\Om\hspace{-.05cm})}(0,0)\rng\ .
\label{yukdef}
\ee
They vanish, unless the three point selection rule is satisfied \cite{ejss}:
$$(1-\Th^{k^\Th}\Om^{k^\Om})f_a+\Th^{k^\Th}\Om^{k^\Om}(1-\Th^{l^\Th}\Om^{l^\Om})f_b
-(1-\Th^{k^\Th+l^\Th}\Om^{k^\Om+l^\Om)})f_c$$
\be
=(1-\Th^{k^\Th}\Om^{k^\Om})\tau_a+\Th^{k^\Th}\Om^{k^\Om}(1-\Th^{l^\Th}\Om^{l^\Om})\tau_b=(1-\Xi)\tau\ .
\label{srule3}
\ee
This equation means that there must be lattice vectors $\tau_a, \tau_b$,
$\tau$ which satisfy it for a given choice of fixed points and sector
numbers. As before, we do the whole factorization independently
in every subspace. Proceeding
%(again $k_j=[k^\Th\al_j^\Th+k^\Om \al_j^\Om],\ l_j=[l^\Th\al_j^\Th+l^\Om
%%\al_j^\Om]$)
as in \cite{ejss} we arrive at the final result for the \z orbifold Yukawa
coupling
\be
\ds{Y_{f_a,f_b,f_c}^{(k^\Th
\hspace{-.22cm},\hspace{.09cm}k^\Om\hspace{-.22cm},\hspace{.13cm}l^\Th
\hspace{-.22cm},\hspace{.09cm}l^\Om)}=
(2 \pi)^{\frac{d}{4}}\lf[ V_{\Lambda}
\prod_{j=1 \atop j \neq \Im}^{d/2} \Ga_{k_j,l_j}\ri]^{1\over2}
\sum_{v \in {\cal U_\bot}}  e^{-\pi v^t({\bf 1}_d+2B)Lv}\ ,}
\label{sss}
\ee
where
\be \ba{cll}
{\cal U} &=& \ds{f_{ba} - \tau_{ba} +
\frac{1-\Th^{k^\Th+l^\Th}\Om^{k^\Om+l^\Om}}{ 1-\Xi}\Lambda\ ,}\nnn
\ds{\Ga_{k_j,l_j}} &=& \left\{ \ba{c@{\quad:\quad}l}
{\ds{\Ga(1-k_j)\Ga(1-l_j)\Ga(k_j+l_j)}
\over\ds{\Ga(k_j)\Ga(l_j)\Ga(1-k_j-l_j)}} &
 \ds{0<k_j+l_j<1} \\
 \noalign{\vspace{0.3cm}}
\ds{\Ga_{1-k_j,1-l_j}} & \ds{1<k_j+l_j<2\ ,}\ea\right.
\ea \label{abbrev}
\ee
and $L$ is a block diagonal matrix whose $j$-th diagonal
block is given by

\be
\ba{clccccr}
    L_j &=&  {\ds{1}\over\ds{|\lambda_j|}}
   \lf[ {\bf 1}_2+ i\,{\rm sgn}(\lambda_j\ri )\epsilon] &,& \epsilon
    = \lf( \ba{clcr} 0 & 1 \nnn -1 & 0 \ea \ri) \, ,  \nnn
    \ds{\lambda_j} &:=&
    \ds{\cot(\pi k_j)+\cot(\pi l_j)}
    &,& \ds{j=1,\ldots,d/2\ .}
\ea
\ee
As in \cite{sjls} one can show that the summation range ${\cal U}$ does not
depend on the particular choice of
$\tau_a$ and $\tau_b$.

\sect{Duality invariance}

The results given above may be used to find the action of duality on the
twisted
sector ground states and to show that the full theory is duality invariant. By
duality we mean here modular duality, i.e. a symmetry acting in the orbifold
moduli space. Since for orbifolds in more than two dimensions the generators of
the duality group are in general not known, we will
focus our attention on the class of
duality transformations described in \cite{MS1,MS2}. These are appropriately
chosen compositions of the "canonical" duality transformation of toroidal
compactifications and $GL(d,\Z)$ transformations.

We will work
in the lattice basis in which the metric and the antisymmetric tensor
field read $g=e^tGe,\ b=e^tBe$, respectively. The twists $\Th$ and
$\Om$ both have to be
 of the lattice $\Lambda$. Therefore we have $\Th e=e Q_1$ and
$\Om
e =e Q_2$, where $Q_1,Q_2$ are integer matrices.
The consistency conditions for \req{action} and \req{bocon} mentioned in
section
2 read now

\bea  Q_1^t(g+b) Q_1 = (g+b) \nnn
 Q_2^t(g+b) Q_2 = (g+b)\ .
\label{modul}
\eea
Similarly to the case of $Z_N$ orbifolds \cite{MS1,MS2}, there is a family of
duality transformations
of \z orbifolds
parameterized by $GL(d,\Z)$ matrices $W$ satisfying ($Q^{\ast} :=
(Q^t)^{-1}$)

\bea
 W\,Q_1\,=\,Q_1^{\ast} W  \nnn
W\,Q_2\,=\,Q_2^{\ast} W\ .
\eea
These duality symmetries act on the background \cite{MS1} by
\bea
\ds{S(g)} &=&  \ds{W \frac{1}{g- b} g \frac{1}{g - b} W^t\ ,} \nnn
\ds{S(b)} &=& \ds{- W \frac{1}{g - b} b \frac{1}{g - b} W^t\ ,}\ \nnn
 \ds{S(e)} &=& e{\ds{1}\over\ds{g-b}}W^t \ .
\label{wdu}
\eea
They are modular (i.e. $S(g+b)$ satisfies \req{modul}) and they generate a
large subgroup of the full modular group.

As in the previous papers \cite{LMN1,LMN2,ejss} we
define the action of duality on the twist fields in such a way that
the twist-anti-twist annihilation couplings \req{svs} remain invariant. If
one accompanies \req{wdu} with

\be
{\sigma}^{k^\Th \hspace{-.22cm},\hspace{.09cm}k^\Om}_{\f_{q}} \mapsto
\tilde{\sigma}^{k^\Th
\hspace{-.22cm},\hspace{.09cm}k^\Om}_{\f_{q}}=\frac{1}{\sqrt{N^\bot_{k^\Th
\hspace{-.22cm},\hspace{.09cm}k^\Om}}}
    \prod_{j=1 \atop j \neq \Im}^{d/2}
\lf[\frac{{\det_j} (GZ-iB)} {{\det_j}(GZ+iB)}\ri]^{\pm \frac{1}{4}
(1-2k_j)} \sum_{a=1}^{N^\bot_{k^\Th \hspace{-.22cm},\hspace{.09cm}k^\Om}} e^{2
\pi i \f_a^t(1-Q_1^{\ast \pm k^\Th}Q_2^{\ast
\pm k^\Om})W\f_q} {\sigma}^{k^\Th
\hspace{-.22cm},\hspace{.09cm}k^\Om}_{\f_{a}}\ ,
\label{tdu}
\ee
where $N^\bot_{k^\Th \hspace{-.22cm},\hspace{.09cm}k^\Om}=
\det_\bot(1-Q_1^{k^\Th}Q_2^{k^\Om})$, then one can prove \req{svs} to be
invariant. The factor in front of
the sum in \req{tdu} is determined by requiring the invariance of the Yukawa
coupling \req{sss} under \req{wdu} and \req{tdu}. For details we refer the
reader to
\cite{ejss}, where this was done for $Z_N$ orbifolds. These calculations can be
generalized to the above cases. Therefore
all the correlation functions calculated in the previous sections are
invariant under the combined duality transformations \req{wdu} and \req{tdu}.

\vskip 2cm
\centerline{\bf Conclusion}
\vskip 0.5cm

We derived the Yukawa couplings of all \z orbifolds in $d$ dimensions. For
our calculation it
was necessary to split
the target space into $d/2$ two dimensional complex subspaces. In these
subspaces we could reduce the calculations to $Z_M$ orbifolds.

One could perform a coordinate transformation into the complex basis to
recover
the dependence of the couplings on the $d/2$ untwisted $(1,1)$--moduli $T_j$
and untwisted
$(2,1)$--moduli $U_i$. Since in general there are fixed planes, not all these
$T_j$ and $U_i$ will
give a contribution in the moduli dependence.

In
particular, for $d=3$ we got couplings depending on the moduli of the
six dimensional orbifold compactifications \cite{FIQ}. They should agree at
their
critical points with the
couplings derived for the corresponding $N=2$ {\em tensor product models}
\cite{G,FIQ2}.

\vskip 1.5cm
\centerline{\bf Acknowledgements}
\vskip 0.5cm

I would like to thank Eung Jin Chun, D. Jungnickel, Hans--Peter Nilles and
Micha\l\  Spali\'{n}ski for
helpful comments.

\end{document}